\documentclass[aps,pra,twocolumn,floatfix,10pt]{revtex4-2}
\usepackage[utf8]{inputenc}
\usepackage[english]{babel}
\usepackage{graphicx}
\usepackage{amsmath,times}
\usepackage{amssymb}
\usepackage{amsfonts}
\usepackage{mathrsfs}
\usepackage{physics}
\usepackage[colorlinks=true,hyperfootnotes=true,breaklinks=true,citecolor=red,urlcolor=blue,linkcolor=blue]{hyperref}
\usepackage{mathbbol} 
\usepackage{forest}
\usepackage{tikz}
\usetikzlibrary{positioning}
\begin{document}
\title{Generalized phase estimation in noisy quantum gates}
\author{Giovanni Ragazzi}
\email{giovanni.ragazzi@unimore.it}
\affiliation{Dipartimento di Scienze Fisiche, Informatiche e Matematiche, 
Universit\`{a} di Modena e Reggio Emilia, I-41125 Modena, Italy}
\author{Simone Cavazzoni}
\email{simone.cavazzoni@unimore.it}
\affiliation{Dipartimento di Scienze Fisiche, Informatiche e Matematiche,  Universit\`{a} di Modena e Reggio Emilia, I-41125 Modena, Italy}
\author{Paolo Bordone}
\email{paolo.bordone@unimore.it}
\affiliation{Dipartimento di Scienze Fisiche, Informatiche e Matematiche, Universit\`{a} di Modena e Reggio Emilia, I-41125 Modena, Italy}
\affiliation{Centro S3, CNR-Istituto di Nanoscienze, I-41125 Modena, Italy}
\author{Matteo G. A. Paris}
\email{matteo.paris@fisica.unimi.it}
\affiliation{Quantum Technology Lab, Dipartimento di Fisica {\em Aldo Pontremoli}, Universit\`{a} degli Studi di Milano, I-20133 Milano, Italy}
\affiliation{INFN, Sezione di Milano, I-20133 Milano, Italy}
\date{\today}
\begin{abstract} 
We examine metrological scenarios where the parameter of interest is encoded onto a quantum 
state through the action of a noisy quantum gate and investigate the ultimate bound to precision 
by analyzing the behaviour of  the Quantum Fisher Information (QFI).  We focus on qubit gates 
and consider the possibility of employing successive applications of the gate. We go beyond the 
trivial case of unitary gates and characterize the robustness of the metrological procedure 
introducing noise in the performed quantum operation, looking at how this affects the QFI 
at different steps (gate applications). We model the dephasing and tilting noise affecting 
qubit rotations as classical fluctuations governed by a Von Mises-Fisher distribution. 
Compared to the noiseless case, in which the QFI grows quadratically with the number of steps, 
we observe a non monotonic behavior, and the appearance of a maximum in the QFI, which defines 
the ideal number of steps that should be performed in order to precisely characterize 
the action of the gate.
\end{abstract}
\maketitle
\section{Introduction}
Quantum Estimation Theory \cite{helstrom1969quantum} is {the mathematical} framework providing 
the optimal {methodologies} to asses the precision in the estimation of one or more parameters 
encoded in the state {of} a quantum system. 
%
%
Phase Estimation \cite{PRXQuantum.2.040301}, in particular, is attracting great attention because of its importance 
not only in metrology \cite{giovannetti2006quantum,cronin2009optics,pitkin2011gravitational} but also 
in quantum information processing in general \cite{shor1994algorithms,russo2021evaluating,PRXQuantum.2.020317,o2019quantum}. Here we consider 
the case in which the parameter to be estimated is encoded in the transformation associated to 
a quantum gate acting on a qubit \cite{PhysRevLett.118.190502,qute20}. The typical {metrological} procedure is to prepare an initial 
state independent from the parameter, apply the quantum gate (one or many times) and then perform 
a measurement to extract information about the {unknown} value of the parameter(s). 

In the ideal case of unitary 
gate operation, the metrological procedure may be easily optimized \cite{dorner2009optimal}, whereas in the presence of noise the problem is not trivial and has not been fully addressed so far.  In turn, the current generation of noisy intermediate-scale quantum devices (NISQ), and likely those in the near future, \cite{cheng2023noisy,preskill2018quantum} are unavoidably affected by noise. The noise affecting NISQ may be classified in different categories
\cite{georgopoulos2021modeling}, including gate infidelities \cite{sutherland2022one}, state preparation and 
measurement (SPAM) errors \cite{yu2023efficient}, decoherence resulting from the interaction with the environment,\cite{shnirman2002noise} and dephasing of the physical qubits. Gate imperfections may be modeled as generalized depolarizing channels \cite{PhysRevA.94.022334,chapeau2024simulating}, i.e., assigning probabilities to the operations described by the Pauli X,Y and Z gates. Another approach is to forget any information about the gate and set an estimation procedure in which this is not required\cite{song2024agnostic}. In this case, doubling the dimension of the probe system and using an an entangled initial state, it is possible to retrieve a relevant amount of information about the parameter \cite{d2001using,gillard2017estimation,rossi2015entangled}.

In this paper, we address realistic noise affecting the transformation performed by the gate, regardless of the causes. In particular, the actual transformation could be different from the one we wanted to apply, and we model the noisy transformation as a weighted sum over all the possible transformations \cite{benedetti2014characterization,rossi2015entangled,ferraro2022universal}. Since every unitary transformation on a qubit can be seen as a rotation of its Bloch vector, we consider two types of noise: one affecting the angle of rotation, referred to as {\em dephasing noise} \cite{PhysRevA.81.012305,teklu2009bayesian}, and the other one affecting the rotation axis, referred to as {\em tilting noise}. The transformations obtained through this procedure cease to be unitary, and we analyze how this affects the estimation 
of {the} gate phase parameter compared to the ideal case.  In order to quantify the precision attainable in the estimation of a single parameter $\theta$, a lower bound is provided by the quantum Crame\'r-Rao theorem, stating
\begin{equation}
\label{Q-CramerRao}
    \hbox{Var}(\theta) \geq \frac 1{M H_{\theta}}
\end{equation}
where $M$ is the number of measurements available and $H_{\theta}$ is a quantity called Quantum Fisher Information (QFI). This tells us that the precision can be arbitrarily increased by increasing the number of measuremets. However, 
this comes with a cost, and in order to exploit each measurement as much as possible we should look at the 
conditions maximizing $H_{\theta}$. In the following, we analyze in details how the dephasing and tilting noise affect the QFI of the phase parameter encoded on the qubit by the gate.

The paper is structured {as follow. In Sec.\ref{sec:TF}, we review the mathematical tools needed for the characterization of the noisy quantum gates. In Sec.\ref{sec:Model}, we evaluate the temporal evolution of the quantum state undergoing successive gate applications. In Sec.\ref{sec:Results}, we illustrate the results obtained for the two types of 
noise, and seek for the optimal number of gate applications that enhances the metrological performance of 
our system. Finally, Sec.\ref{sec:Summary} closes the paper with some concluding remarks. Additional 
material can be found in Appendices \ref{app:NoLoss}-\ref{app:NoisyMatrices}.}

\section{Metrology on the Bloch sphere}
\label{sec:TF}
The most general qubit state may be written as 
\begin{equation}
    \vert \psi \rangle = \cos \alpha \vert 0 \rangle + e^{i\gamma} \sin \alpha \vert 1 \rangle \quad 0\leq \alpha,\frac{\gamma}2 \leq \pi\,,
\end{equation}
and may be associated to a vector on the unit sphere, 
usually referred to as Bloch vector: $\vec b= \left(\sin\alpha\cos\gamma,\sin{\alpha}\sin{\gamma},\cos{\alpha}\right)^T$. For mixed states $\rho$, the 
Bloch vector is obtained by
\begin{equation}
    \vec b = \Tr{\rho\, \vec{\sigma}}
\end{equation}
where $\vec{\sigma}$ is the vector of Pauli matrices:
\begin{equation}
    \vec{\sigma}=\left[ \left(\begin{array}{cc}
       0  &  1\\
         1& 0
    \end{array}\right), \left(\begin{array}{cc}
       0  &  -i\\
         i& 0
    \end{array}\right),\left(\begin{array}{cc}
       1  &  0\\
         0& -1
    \end{array}\right) \right]^T.
\end{equation}
A quantum gate encodes a transformation performed on a quantum state. To map pure states to pure states (as the evolution of a quantum isolated state should do) the corresponding matrix must be unitary: $U \in U(2)$ since we are acting on a qubit. However, we will consider only special unitary matrices: $U \in SU(2)$, because every unitary matrix can be written as a special unitary times a phase, whose action is trivial on a quantum state. 
A unitary transformation acting on a qubit corresponds to a rotation of its Bloch vector. To see this we write $U$  by exponentiation of the $SU(2)$ generators: $U=e^{-\frac{i}{2}\,\theta \hat{n} \cdot \vec{\sigma}}$ with $\vec{\theta}=\theta \hat n$ being the generalized angles \cite{cacciatori2022compact}. It transforms a  state $\rho_0$ as $\rho=U\rho_0U^{\dagger}$. Taking first order of an infinitesimal rotation 
\begin{align}
\label{State_infinitesimal_rotation}
    \rho&= e^{-\frac{i}{2}\,\theta \hat{n} \cdot \vec{\sigma}}\, \rho_0\, e^{\frac{i}{2}\,\theta \hat{n} \cdot \vec{\sigma}} \simeq \rho_0-i\frac{\theta}2 \left[ \vec{\sigma} \cdot \hat n , \rho_0 \right]\,,
\end{align}
the corresponding Bloch vector is transformed as
\begin{align}
\label{Bloch_infinitesimal_rotation}
    \vec{b}&=\Tr{\rho\, \vec{\sigma}} \simeq\vec b_0-i\frac{\theta}2 \Tr{\left[\vec{\sigma} \cdot \hat{n}, \rho_0\right]\vec{\sigma}} \notag \\
    &=\vec b_0+ \theta\, \hat n \cross \vec b_0
\end{align}
so that the (special) unitary transformation $e^{-\frac{i}{2}\,\theta \hat{n} \cdot \vec{\sigma}}$ corresponds to the rotation $R_{\hat n}( \theta)$,  around the axis $\hat n$. 

In parameter estimation, the goal is to estimate the value of a parameter $\theta$ by sampling a stochastic variable $X$, whose conditional distribution $p(x\vert\theta)$ is the probability to obtain outcome $x$ provided that the the parameter has value $\theta$.
A fundamental result is the Cram\'er-Rao inequality \cite{cramer1999mathematical,rao1947minimum}, providing a lower bound on the variance of $\theta$ when it is by measuring the variable $X$:
\begin{equation}
\label{C-CramerRao}
    \hbox{Var}(\theta) \geq \frac 1{M F_x(\theta)}.
\end{equation}
The difference with Eq.(\ref{Q-CramerRao}) is that in the right hand side we have $F_x(\theta)$, termed Fisher Information (FI), defined as
\begin{equation}
\label{FI}
F_x(\theta)=\sum_x\frac{\left[\partial_{\theta}p(x\vert\theta)\right]^2}{p(x\vert\theta)}\,.
\end{equation}
 In a quantum mechanical framework $\theta$ would be a parameter of a quantum state. Measurements are performed through positive operators (POVM) $\left\{\Pi_x\right\}$ with $\sum_x \Pi_x=\mathbb 1$, and can be seen as tools to generate the probability distributions $p(x\vert\theta)$. In this way, some measurements will generate probability distributions with a high FI while others will be less efficient. By maximizing the FI with respect of all the possible measurements one obtains the the quantum Cram\'er-Rao  bound \cite{paris2009quantum} of Eq.(\ref{Q-CramerRao}), having then 
\begin{equation}
\label{ineqQF}
    H_\theta \geq F_x(\theta).
\end{equation}
In this way, the QFI depends only on the quantum state $\rho_\theta$ and is defined as
\begin{equation}
    \label{QFI}
    H_\theta= \max_x \{F_x(\theta)\}=\Tr{\rho_\theta \mathcal L^2_\theta},
\end{equation}
where $\mathcal L_{\theta}$ is an hermitian operator, referred to as the Symmetric Logarithmic Derivative (SLD). It is related to the rate at which the state changes when $\theta$ varies and is implicitly defined as
\begin{equation}  \partial_{\theta}\rho_{\theta}=\frac{\mathcal{L}_{\theta}\rho_{\theta}+\rho_{\theta}\mathcal{L}_{\theta}}2.
\end{equation}
{Evaluating the QFI for a quantum state gives an upper bound on the precision, which may be achieved by measuring  the projectors on the eigenstates of the SLD operator.}

The evaluation of $H_{\theta}$ usually involves the diagonalization of the state density matrix. For qubits this leads to \cite{zhong2013fisher,chapeau2015optimized} 
\begin{equation}
\label{BlochQFI}
    H_{\theta}=\left\{ \begin{array}{lr}
        \vert\partial_{\theta}\vec b_{\theta}\vert^2+\frac{(\vec b_{\theta} \cdot \partial_{\theta}\vec b_{\theta})^2}{1-\vert \vec b_{\theta}\vert^2} &  \mbox{if} \, \vert \vec b_{\theta}\vert<1\\
        & \\
        \vert\partial_{\theta}\vec b_{\theta}\vert^2 & \mbox{if} \, \vert\vec b_{\theta}\vert=1
    \end{array}\right. ,
\end{equation}
where $\vec b_{\theta}$ is the Bloch vector associated to the state $\rho_{\theta}$.

The Cram\'er-Rao theorem is valid for those values of the parameter where 
the QFI is continuous, otherwise the bound does not hold \cite{vsafranek2017discontinuities,seveso2019discontinuity}. In the quantum case, discountinuities usually occur at values of the parameter for which the rank of the density matrix changes. In our case, this happens for $\theta=0$, and  therefore we consider the domain $0<\theta<2\pi$. This happens because for $\theta=0$ the rotation around any axis is the identity and is left invariant by any form of noise, while for $\theta\neq0$ fluctuations of $\theta$ turn any unitary rotation into a purity decreasing quantum operation.
 
{Since we want to characterize a noisy system where the parameter is the 
rotation amplitude $\theta$ around a given rotation axis, the fluctuations 
of those parameters happen on a circular or spherical domain. A convenient distribution is thus given by the  von Mises-Fisher probability density function \cite{kurz2016kullback}, defined as
\begin{equation}
    p_k^d(\hat x)= C_d(k) e^{k \hat x \cdot \hat{\mu}}
\end{equation}
where $\hat x, \hat{\mu}$ are vectors belonging to the $d-1$ sphere $S^{d-1}$ and $C_d(k)$ is a normalization factor that depends on the dimension
\begin{equation}
    C_d(k)=\left(\int_{S^{d-1}} e^{k \hat x \cdot \hat{\mu}} d\hat x\right)^{-1}=\frac{k^{\frac d2-1}}{(2\pi)^{\frac d2} I_{\frac d2-1}(k)}
\end{equation}
with $I_n(k)$ denoting the modified Bessel function of the first kind of order $n$ evaluated at $k$. The quantity $\hat{\mu}$ identifies the predominant direction on the hypersphere while $k$ represents a concentration parameter, and controls how fast the probability decreases when $\hat x$ deviates from $\hat{\mu}$.

For $d=2$ we are considering a stochastic variable $\epsilon$ living on a circle and the corresponding density function, known as von Mises distribution, 
reduces to
\begin{equation}
    p_k^{d=2}(\epsilon)=\frac{e^{k \cos(\epsilon-\mu)}}{2\pi I_0(k)}.
\end{equation}
An alternative expression in series of cosines is given by
\begin{equation}
\label{VMExpansion}
    p_k^{d=2}(\epsilon)=\frac1{2\pi} \left\{1+\frac2{I_0(k)} \sum_{j=1}^{\infty} I_j(k) \cos [j(\epsilon-\mu)]\right\}.
\end{equation}
The $d=3$ case was introduced \cite{fisher1953dispersion} to provide the correct distribution for elementary errors over the surface of the unit sphere, pointing out that the most used Gaussian distribution loses validity when the entity of the errors forces to consider the actual topology. The probability density reads
\begin{equation}
    p^{d=3}_{k}(\hat x)=\frac{k e^{k \hat x \cdot \hat{\mu}}}{4\pi \sinh(k)}.
\end{equation}

\section{The statistical model}
\label{sec:Model}
We address a parameter estimation problem in which the dependence on the 
parameter $\theta$ is encoded onto a quantum state by the application of 
a $\theta$-dependent quantum gate. In particular, we consider the scenario in which the quantum gate, due to imperfections leading to fluctuations, instead of performing the desired transformation may perform a different one. Let us denote by  $U_{\theta}$ the intended transformation and $U_{\boldsymbol{\lambda},\theta}$ the possible transformations that are actually performed. Here $\boldsymbol{\lambda}$ denotes a $d$-dimensional parameter ($d \leq 3$) so that $U_{\boldsymbol{\lambda},\theta}$ runs over a certain subset of the $SU(2)$ matrices.  The gate performs the transformation 
$U_{\boldsymbol{\lambda},\theta}$ with probability
$p(\boldsymbol{\lambda})$, with $\int{p(\boldsymbol{\lambda}) d\boldsymbol{\lambda}}=1$. If $\rho_0$ is the the initial state of the qubit, after $t$ gate iterations, we are left with the state
\begin{align}
    \rho_t= \int...\int &p(\boldsymbol{\lambda}_1)...p(\boldsymbol{\lambda}_t) U_{\boldsymbol{\lambda}_t,\theta}...U_{\boldsymbol{\lambda}_1,\theta} \nonumber \\ &\cross \rho_0 U_{\boldsymbol{\lambda}_1,\theta}^{\dagger}...U_{\boldsymbol{\lambda}_t,\theta}^{\dagger} d\boldsymbol{\lambda}_1...d\boldsymbol{\lambda}_t.
\end{align}
The quantity $t$ is discrete and denotes the number of gate iterations. We nevertheless refer to $t$ as the 'time'. The  Bloch vector at time $t$ is given by 
\begin{align}
    \vec b_t&=\Tr{\rho_t\, \vec{\sigma}} \nonumber \\
    &= \int\!...\!\int p(\boldsymbol{\lambda}_1) ... p(\boldsymbol{\lambda}_t) \Tr \bigl\{ U_{\boldsymbol{\lambda}_t,\theta} ... U_{\boldsymbol{\lambda}_1,\theta} \nonumber \\ 
    &\quad \quad \quad \quad \cross \rho_0 U_{\boldsymbol{\lambda}_1,\theta}^{\dagger}...U_{\boldsymbol{\lambda}_t,\theta}^{\dagger} 
    \, \vec{\sigma}
    \bigr\} d\boldsymbol{\lambda}_1 ... d\boldsymbol{\lambda}_t \nonumber \\
    &= \int\!...\!\int p(\boldsymbol{\lambda}_1) ... p(\boldsymbol{\lambda}_t) \left[ R_{\boldsymbol{\lambda}_1}\!(\theta) ... R_{\boldsymbol{\lambda}_t}\!(\theta)\,\vec b_0 \right] d\boldsymbol{\lambda}_1 ...d\boldsymbol{\lambda}_t
\end{align}
where 
$R_{\boldsymbol{\lambda}}(\theta)$ 
is the Bloch rotation corresponding to $U_{\boldsymbol{\lambda},\theta}$. If we assume that the probability distribution $p(\boldsymbol{\lambda})$ is the same at each gate iteration and is independent on the 
state on which the gate acts, linearity allows us to write 
\begin{align}
    \vec b_t&= 
    \int p(\boldsymbol{\lambda}_t)R_{\boldsymbol{\lambda}_t}\!(\theta)d\boldsymbol{\lambda}_t ... 
    \int p(\boldsymbol{\lambda}_1)R_{\boldsymbol{\lambda}_1}\!(\theta)d\boldsymbol{\lambda}_1
    \, \vec b_0 \\
    &= \left[\Tilde{R}_k(\theta)\right]^t \vec b_0\,.
\end{align}
This tells us that once we compute the `noisy rotation'
\begin{equation}
\label{NoisyRotationGeneral}
    \Tilde{R}_k (\theta)\equiv \int\! d\boldsymbol{\lambda}\,p(\boldsymbol{\lambda})\,
    R_{\boldsymbol{\lambda}}(\theta)\,,
\end{equation}
 we can simply apply it to the initial Bloch vector $\vec b_0$ to obtain the evolved one $\vec b_t$ at each gate iteration. Alternatively, we may also use the following expressions \cite{cavazzoni2024coin}
 \begin{align}
    \vec b_t(\theta,k)&=\left[\Tilde{R}_{k}(\theta)\right]^t \vec b_0\\
     \partial_{\theta}\vec b_t(\theta,k)&=\partial_{\theta}\left[\Tilde{R}_{k}(\theta)\right]^t \vec b_0\,,
 \end{align}
where
\begin{align}
    \partial_{\theta}\left[\Tilde{R}_{k}(\theta)\right]^t= 
    \sum_{j=0}^t\left[\Tilde{R}_{k}(\theta)\right]^j 
    \partial_\theta\Tilde{R}_{k}(\theta) \left[\Tilde{R}_{k}(\theta)\right]^{t-j-i}.
\end{align}
Once we know $\vec b_t(\theta,k)$ and  $\partial_{\theta}\vec b_t(\theta,k)$ we can compute the QFI at time $t$ . 

\section{Results}
\label{sec:Results}
In this Section, we illustrate results. We start by considering systems affected by {\em pure dephasing}, i.e., noise affecting only the angle of rotation, and then examine the case of {\em pure tilting}, where noise is  affecting only the axis of rotation. Finally, we consider the general case where both kind of noise are present. Notice that this is the most general kind of noise that may occur for a qubit rotation, which is characterized by the $3$ degrees of freedom of a 3-dimensional rotation (one for dephasing, and two for tilting). As we will see in details, the presence of noise dramatically alters the behavior of the QFI compared to the noiseless case. The most striking difference being that the QFI does no longer increase indefinitely with the number of gate applications, but instead reaches a maximum and then goes to zero.
\subsection{Noiseless evolution}
Let us start with the noiseless case, in which the gate performs exactly the desired operation \cite{liu2015quantum}. Starting from a generic state, we have 
\begin{equation}
    \vec b_0=\vert \vec b_0\vert \left( \sin{\alpha}\cos{\gamma}, \sin{\alpha}\sin{\gamma}, \cos \alpha \right)^T\,.
\end{equation}
We apply $t$ times a $\hat z$ rotation of an angle $\theta$, that is the parameter to be estimated. We choose a $\hat z$ rotation without loss of generality (see Appendix \ref{app:NoLoss})
\begin{equation}
    R_{\hat z} (\theta)=\left(\begin{array}{ccc} 
       \cos\theta  & -\sin\theta & 0 \\
        \sin \theta & \cos\theta & 0\\
        0&0&1 
    \end{array}\right).
\end{equation}
and after $t$ iterations the state is 
\begin{align}
    \vec b_t & = \left[R_{\hat z} (\theta)\right]^t \vec b_0
    =R_{\hat z}(t\, \theta) \vec b_0 \nonumber \\
    & =\vert \vec b_0\vert\left(\begin{array}{c}
       \sin{\alpha}\cos{(t\, \theta+\gamma)}  \\
        \sin{\alpha}\sin{(t\, \theta+\gamma)} \\
        \cos \alpha
    \end{array} \right) \,.
\end{align}
Using Eq.(\ref{BlochQFI}) it is easy to obtain
\begin{align}
\label{QFI_no_noise}
    H_{\theta}(t)=\vert \partial_{\theta}\vec b_t\vert^2=\vert \vec b_0\vert^2 t^2 \sin^2 \alpha .
\end{align}
The QFI increases with the purity of the initial state, which is not spoiled by the simple 
rotation applied. Indeed, as observed in \cite{chapeau2015optimized}, the QFI always 
increases with the purity of the measured qubit state.  
In addition, we see that the QFI grows quadratically with the number of gate iterations, and depends on the polar angle $\alpha$ of the initial state through a factor $\sin^2\alpha$. This is in accordance with a well known result \cite{toth2018quantum}, stating that when the dependence on a parameter is given to a quantum state through the action of a unitary transformation $U_{\theta}=e^{-i\theta T}$, the QFI is maximized by taking the initial state in the form $\vert\psi_0\rangle=\frac1{\sqrt2}\left(\vert e_{max}\rangle+\vert e_{min}\rangle \right)$, where $\vert e_{max}\rangle$ and $\vert e_{min}\rangle$ are the eigenstates of the generator of the transformation $T$ associated to the highest and lowest eigenvalues. Here the generator would be proportional to the Pauli matrix $\sigma_z$, whose eigenstates are the elements of the trivial basis $\vert g \rangle$ and $\vert e \rangle$. An equal superposition of them would lie on the equator of the Bloch sphere, having then $\alpha=\frac{\pi}2$. Interestingly, the QFI does not depend from the actual rotation angle $\theta$ that we want to estimate.
\subsection{Pure Dephasing}
{After the analysis of the noiseless evolution we move to the study of the effect of the different noise models.} In this Section, we assume that the noise is affecting only the rotation, while the axis of rotation is preserved. In other words, we assume that the gate instead of performing the rotation $R_{\hat z}(\theta)$, may perform the rotation $R_{\hat z}(\theta+\epsilon)$ with probability $$p_{k_D}(\epsilon)=\frac{e^{k_D\cos\epsilon}}{2\pi I_0(k_D)}\,,$$
where $I_j(x)$ denotes the modified Bessel function of the first kind of order $j$.
We consider the axis of rotation to be the 
$\hat z$ direction to simplify the computation, but the results hold general by simply rotating the initial vector (see Appendix \ref{app:NoLoss}). The probability of obtaining a certain rotation is governed by a Von Mises distribution centered in $\epsilon=0$ (i.e., we assume an unbiased gate where the most likely rotation is the nominal one) with concentration parameter $0\leq k_D \leq \infty$. This quantity determines the quality of the quantum gate: for $k_D \rightarrow \infty$ we recover the noiseless result, while for $k_D \rightarrow 0$ the angle of rotation becomes completely random.
\par
The resulting dephased rotation is (see Appendix \ref{app:NoisyMatrices})
\begin{equation}
    \Tilde{R}_{\hat z,k_D}(\theta)=\left(\begin{array}{ccc} 
     \mathcal{F}_{k_D}\cos\theta  & -\mathcal{F}_{k_D}\sin\theta & 0 \\
       \mathcal{F}_{k_D}\sin \theta & \mathcal{F}_{k_D}\cos\theta & 0\\
        0&0&1 
    \end{array}\right)
\end{equation}
where $\mathcal{F}_{k_D}=I_1(k_D)/I_0(k_D)$. The operation $\Tilde{R}_{\hat z,k_D}(\theta)$ is still rotating the qubit, but also shrinking the Bloch vector in the $xy$ plane. {Indeed, the main difference between a pure and a noisy rotation affected by dephasing stays in the coefficient $\mathcal{F}_{k_D}$, which assesses the precision of the operation, according to the limit}
\begin{equation}
    \label{eq:I1/I0}
    \lim_{k_{D} \rightarrow \infty} \mathcal{F}_{k_D} = 1
\end{equation}
The QFI after $t$ gate applications (see Appendix \ref{app:NoisyMatrices}) is given by
\begin{align}
\label{QFIDephasing}
    H_{\theta}(t)=\big(\mathcal{F}_{k_D}\big)^{2t}\, t^2 \vert\, \vec b_0\vert^2 \sin^2 \alpha\,.
\end{align}
The behaviour of $ H_{\theta}(t)$ as a function of the number of gate applications is shown in Fig. \ref{fig:Dephasing_many_k} for different values of the concentration parameter $k_D$.
We see that noise introduces a non-monotonic behavior {depending on the ratio $\frac{I_1(k_D)} {I_0(k_D)}$}, physically motivated by the fact that the gate gives to the state dependence from $\theta$, but also decreases its purity. Knowing this, it is interesting to find the ideal time $t_{ideal}$, namely the one maximizing Eq.(\ref{QFIDephasing}). 
\begin{figure}[h!]
    \centering
    \includegraphics[width= 0.95\columnwidth]{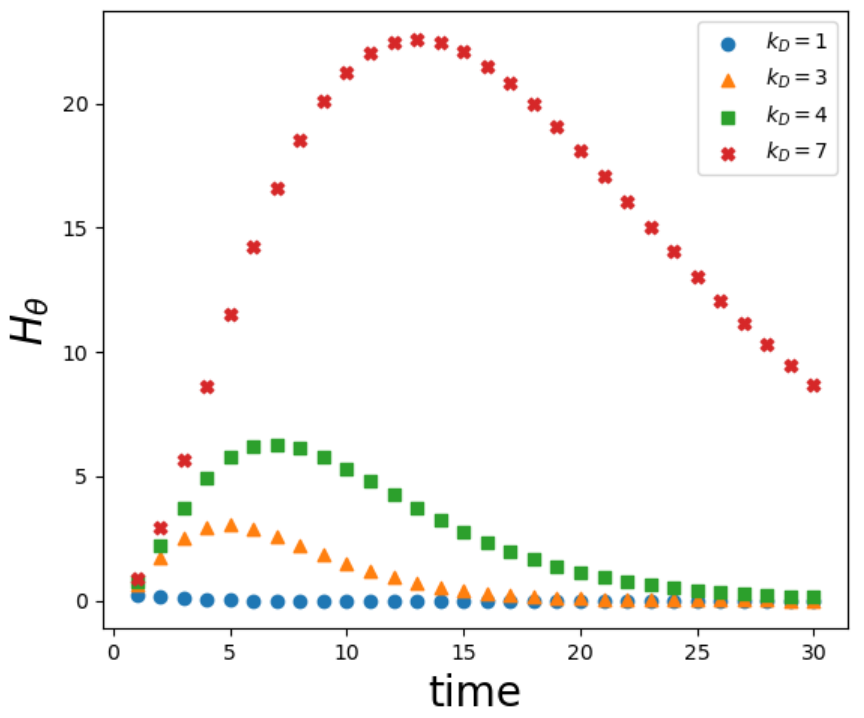}
    \caption{QFI as a function of the  number of applications of a dephased rotation for different values of the concentration parameter $k_D$. The initial state is $\vec b_0=(1,0,0)^T$, the rotation axis is $\hat n=(0,0,1)^T$ and the nominal 
    rotation angle is $\theta=\frac{\pi}4$. The higher is $k_D$ (and so the quality of the quantum gate) the higher is the maximum value of the QFI before it starts decreasing.
    \label{fig:Dephasing_many_k}}
\end{figure}

The optimal value of $t$ is given by 
\begin{equation}
    t_{opt}=-\big(\ln\mathcal{F}_{k_D}\big)^{-1}
\end{equation}
Of course, since it represents a number of gate applications, we have to select the closest integer to $t_{opt}$. Concerning the initial state, the QFI is still maximized taking the starting vector lying on the equator. We see also that, as in the noiseless case, the QFI does not depend on the rotation angle $\theta$.
\subsection{Pure Tilting}
{After the analytical characterization of the dephasing case,} in this Section we address the case where the noise is affecting only the axis of rotation, preserving the rotation angle {$\theta$}. In other words, we assume that the gate, instead of performing the target rotation $R_{\hat z}(\theta)$, is applying the rotation $R_{\hat n}(\theta)$ with a probability density given by the Von Mises-Fisher distribution on the 2-Sphere, i.e., 
$$p_{k_T}(\hat{n}) = \frac{k_T}{4\pi\sinh{k_T}}\,e^{k_T \cos{\phi}}\,,$$ 
where $\hat n= \left(\sin\phi\cos\varphi,\sin{\phi}\sin{\varphi},\cos{\phi}\right)^T$. 
The probability density has concentration parameter $k_T$ and is centered in the direction of the $\hat z$ axis (again without loss of generality, see Appendix \ref{app:NoLoss}). We refer to it as the 
"central axis", because "rotation axis" would be misleading since we are rotating around all the possible ones.
Following the computation in Appendix \ref{app:NoisyMatrices}, one obtains the tilted rotation $\Tilde{R}_{\hat z, k_T}(\theta)$, shown in Eq. (\ref{TiltingMain_app}). As expected, we recover a pure $\hat z$ rotation in the limit $k_T\rightarrow \infty$. On the other hand, for vanishing $k_T$ one obtains 
a transformation proportional to the identity
\begin{equation}\label{rut}
    \Tilde{R}_{UT}(\theta) \xrightarrow[]{k_T\rightarrow 0} \frac{1+2\cos\theta}3 \mathbb{1}
\end{equation}
This case, termed uniform tilting, will be treated later. Except for this case, the general expression 
of $\vec b_t = \left[\Tilde{R}_{\hat z, k_T}(\theta)\right]^t \vec b_0 $ becomes 
cumbersome, as that of the QFI, and will not be reported here. Rather, in Fig. \ref{fig:Tilting_many_k} we show the behaviour of the QFI as a function of the number of applications of a tilted rotation for different values of the concentration parameter $k_T$. As it is apparent from the plot, the behavior is qualitatively similar to the one obtained for dephasing noise: increasing the concentration parameter $k_T$ increases the highest value that is reached by the QFI.
\begin{figure}[h!]
    \centering
    \includegraphics[width= 0.95\columnwidth]{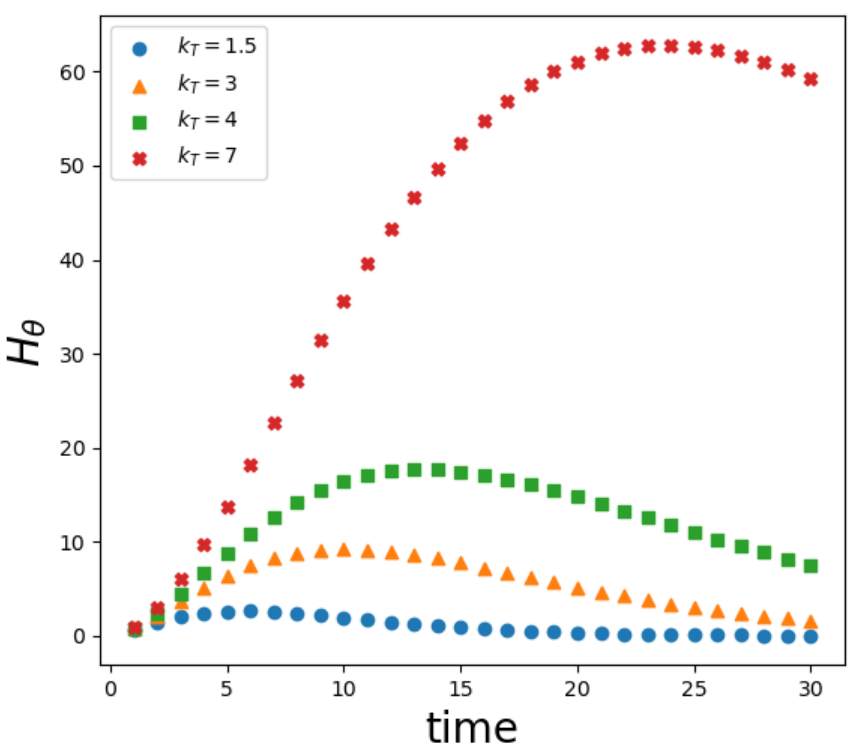}
    \caption{QFI as a function of the number of applications of a tilted rotation for different values of the concentration parameter $k_T$. The initial state is $\vec b_0=(1,0,0)^T$, the central axis is $\hat n=(0,0,1)^T$ and the rotation angle is $\theta=\frac{\pi}4$. Increasing $k_T$ means considering a gate less affected by tilting noise, allowing the QFI to reach higher values.}
    \label{fig:Tilting_many_k}
\end{figure}

However, despite the similarities with the dephasing noise case, tilting noise introduces a dependence in the QFI from the rotation angle $\theta$. In particular, as we can see in Fig. \ref{fig:Tilting_many_theta}, the QFI achieves higher values for small rotation angles, where the effect of tilting noise is apparently reduced.
%
The case in which we have no control at all on the axis of rotation can be treated analytically. 
Indeed, in this case the noisy transformation $ \Tilde{R}_{UT}(\theta)$ is that of Eq. (\ref{rut}) 
which can be obtained computing 
\begin{equation}
     \Tilde{R}_{U.T}(\theta)=\frac1{4\pi}\int_0^{2\pi}\!\!\!d\varphi\int_0^{\pi} \!\!\!d\phi \,\sin\phi\, 
     R_{\hat n}(\theta)\,.
\end{equation}
Correspondingly, one has 
\begin{align}
    \vec b_t=&\left( \frac{1+2\cos\theta}3 \right)^t \vec b_0\\
    \partial_{\theta}\vec b_t=&- \frac23 \sin\theta t \left( \frac{1+2\cos\theta}3 \right)^{t-1} \vec b_0
\end{align}
and using Eq.(\ref{BlochQFI}),
\begin{equation}
\label{UTiltingQFI}
    H_{\theta}^{k_T\rightarrow 0}(t) = \frac49\, t^2 \, \frac{Z^{2t-2}}{1-Z^{2t} \vert \vec b_0\vert^2} \sin^2\theta \vert \vec b_0\vert^2.
\end{equation}
where 
$$Z=\frac13 \big(1+2\cos\theta\big)$$
\begin{figure}[h!]
    \centering
    \includegraphics[width= 0.95\columnwidth]{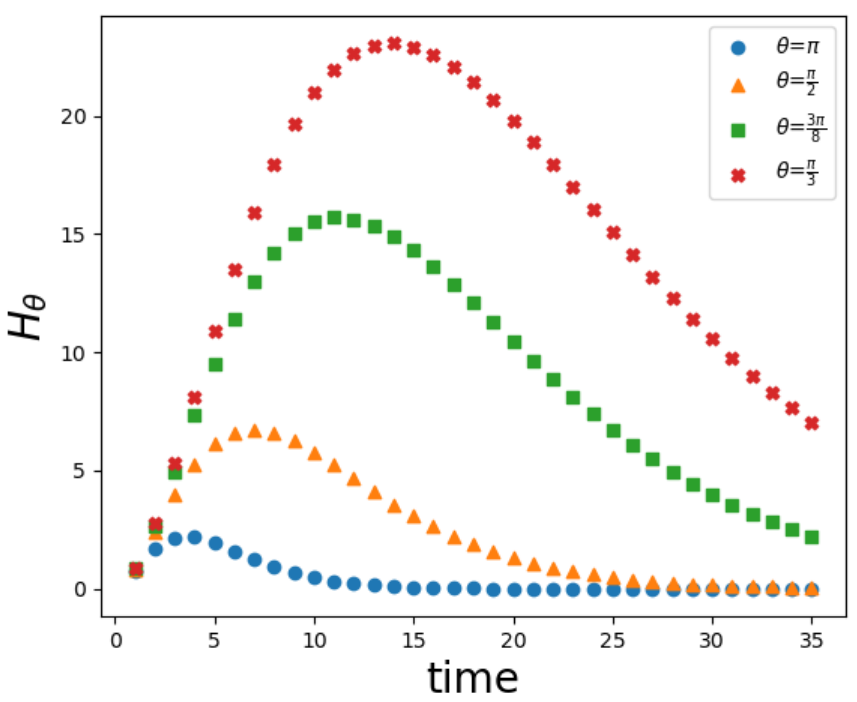}
    \caption{QFI as a function of the number of applications of a tilted rotation for different values of the rotation angle $\theta$. The initial state is $\vec b_0=(1,0,0)^T$, the rotation axis is $\hat n=(0,0,1)^T$ and the concentration parameter is $k_T=7$. By reducing the angle of rotation we can mitigate the effect of the tilting noise. }
    \label{fig:Tilting_many_theta}
\end{figure}

It is interesting to analyze the case in which the initial state is pure ($\vert \vec b_0\vert^2$=1). In this case, if $\theta\rightarrow0$, we obtain a linear dependence on time:
\begin{equation}
    H_{\theta}^{k_T\rightarrow0}(t) \xrightarrow[]{\theta\rightarrow 0} \frac23 t.
\end{equation}
We should note that in this case our model falls into the one addressed in \cite{song2024agnostic}, where they considered one single gate application. Using two entangled qubits they recovered the maximum value of QFI ($H_\theta=1$), while using an entangle-free couple of qubits they set up a measurement procedure resulting in a FI of $\frac23$ regardless of the rotation angle. In our case instead, using a single qubit probe, the QFI in Eq.(\ref{UTiltingQFI}) depends on the rotation angle $\theta$, and for $t=1$ reads
\begin{equation}
    H_{\theta}^{k_T\rightarrow0}(t=1)=\frac{2\cos^2 \frac{\theta}2 }{2+ \cos\theta}\,,
\end{equation}
which assumes its maximum $H_{\theta}^{k_T\rightarrow0}(t=1)=2/3$ for $\theta=0$.

Concerning the dependence from the initial state (in particular from its polar angle $\alpha$), the highest value of the the QFI is still reached selecting an initial vector lying on the equator {of the Bloch sphere} (having  $\alpha=\frac{\pi}2$) as we can see in Fig. \ref{fig:Tilting_many_angles}. However, the dependence from $\alpha$ is not anymore a simple $\sin^2\alpha$ factor.
\begin{figure}[h!]
    \centering
    \includegraphics[width= 0.95\columnwidth]{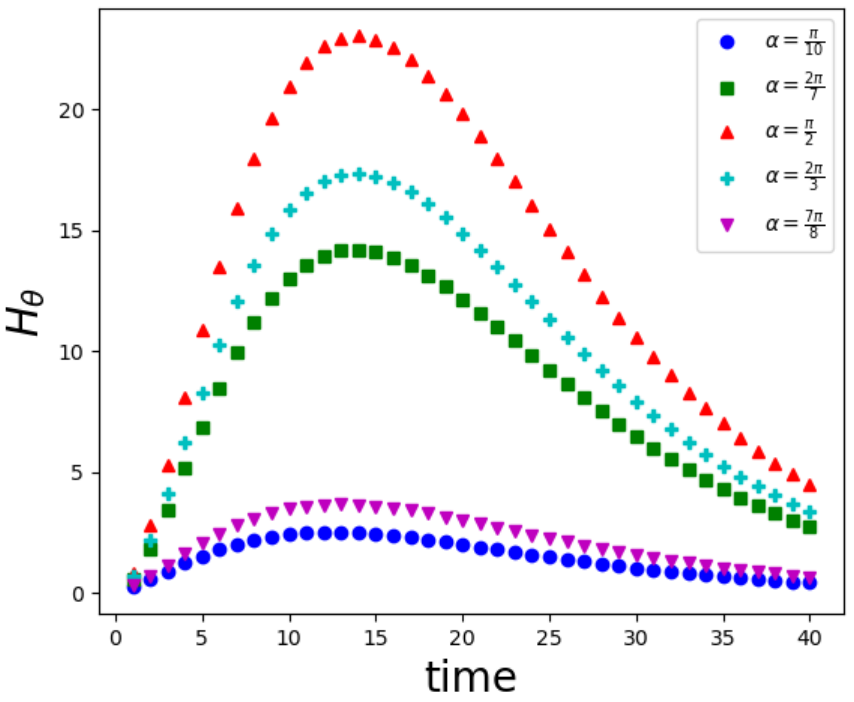}
    \caption{QFI as a function of the number of applications of a tilted rotation for different values of the initial state polar angle $\alpha$. In particular, the initial vectors are chosen in the form $\vec b_0=( \sin\alpha,0,\cos\alpha)^T$. The central axis is $\hat n=(0,0,1)^T$, the rotation angle is $\theta=\frac{\pi}{3}$ and the concentration parameter is $k_T=7$. We can see that the higher curve is the one associated with $\alpha=\frac{\pi}2$.}
    \label{fig:Tilting_many_angles}
\end{figure}

\subsection{General Noise}
As already mentioned, the most general effect of the noise is a combination of both a tilting and dephasing. The gate performs the rotation $R_{\hat n}(\theta+\epsilon)$ instead of the desired $R_{\hat z}(\theta)$, where dephasing noise is responsible for fluctuations in the  rotation angle {uncertainty} and the tilting one for the axis {instability}. The probability governing the fluctuations of the rotation angle is a Von Mises distribution centered in $\epsilon=0$ with concentration parameter $k_D$. The tilting noise is described by a Von Mises-Fisher distribution centered in the central axis $\hat z$, with concentration parameter $k_T$. The noisy rotation $\Tilde{R}_{\hat z, k_D, k_T}(\theta)$ is obtained in \ref{app:NoisyMatrices} and shown in \ref{GeneralMain_app}.

As intuitively expected, in the noiseless limit ($k_D \rightarrow \infty, k_T \rightarrow \infty$) we recover a pure $\hat z$ rotation. Removing the dephasing noise ($k_D\rightarrow \infty$) leaves us with a tilted rotation $\Tilde{R}_{\hat z,k_T}(\theta)$ and thus for $k_T\rightarrow \infty$ we recover a dephased rotation $\Tilde{R}_{\hat z,k_D}(\theta)$. {Contrariwise} , in the limit $k_D\rightarrow 0$, we no longer have a dependence on $\theta$, resulting in a vanishing QFI. For $k_T\rightarrow 0$ while keeping $k_D>0$, {the QFI assumes the form}
\begin{align}
    H_{\theta}^{k_T\rightarrow0}(t) & = \frac49\,t^2 
    \frac{Z^{2t-2}}
    {1-Z^{2t} \vert \vec b_0\vert^2}   \left[\frac{I_1(k_D)}{I_0(k_D)}\sin\theta\right]^2 \vert\vec b_0\vert^2.
\end{align}
where here the quantity $Z$ in Eq. (\ref{UTiltingQFI}) becomes
$$Z= \frac13 \left(1+2\,\frac{I_1(k_D)}{I_0(k_D)}\cos\theta\right)\,.$$
This is analogous to the uniform tilting studied in the previous section, but with additional $\frac{I_1(k_D)}{I_0(k_D)}$ factors which cause 
\begin{equation}
   H_{\theta}^{k_T\rightarrow0}(t) \xrightarrow[]{\theta\rightarrow 0} 0
\end{equation}
unless we perform the limit $k_D\rightarrow\infty$ in advance. In the general case, 
the characterization of $H_{\theta}$ is more challenging, since it depends on several parameters: $t,\theta, k_D, k_T, \vert \vec b_0\vert^2$ and the angle $\alpha$ between the central axis and the initial state. For this reason we assume to start with a pure state $\vert \vec b_0\vert^2=1$ and focus on two interesting cases.
%

\begin{figure}[h!]
    \centering
    \includegraphics[width= 0.95 \columnwidth]{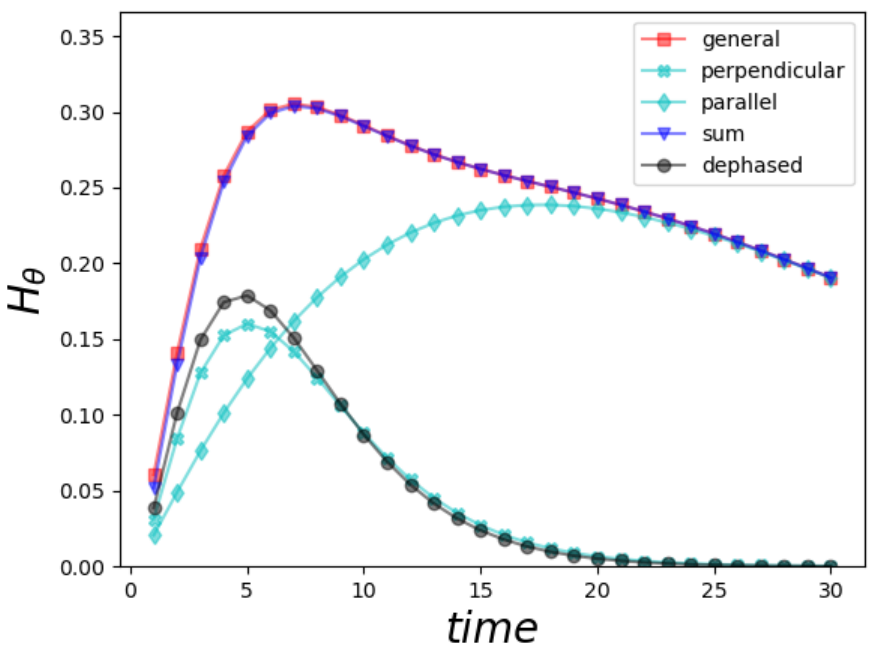}
    \caption{Here we plot the time evolution of different QFIs. The initial state is $\vec b_0=\frac{1}{\sqrt{17}}(0,1,4)^T$ and the central axis is $\hat n=(0,0,1)^T$. The nominal angle of rotation is $\theta=\frac{\pi}8$ and the concentration parameters are $k_D=3$ and $ k_T=10$. Starting from $\vec b_0$, \textit{complete} denotes the QFI obtained with both tilting and dephasing noise while \textit{dephased} is pure dephasing. The curves \textit{perpendicular} and \textit{parallel} are obtained starting from $\vec b_0^{\perp}=\frac{1}{\sqrt{17}}(0,1,0)^T$ and $\vec b_0^{\shortparallel}=\frac{1}{\sqrt{17}}(0,0,4)^T$ respectively and are affected by both the noises. Their sum is plotted as \textit{sum}. We can see that the \textit{perpendicular} curve stays below the curve obtained with pure dephasing (except for a little crossing in the tail) while the \textit{parallel} curve does not have to do so. Their sum provides a good approximation for the real QFI curve, that is much higher than the \textit{dephased} one despite being affected by one additional noise source. }
    \label{fig:decouple}
\end{figure}

{As a first example, we consider an equatorial initial state with $\alpha=\pi/2$, which is optimal for pure dephasing and pure tilting. We observe again a maximum at a certain time and then a slow decrease.  The lower are the noises {effect} (i.e.m the higher their concentration parameters) the higher is the maximum and also the time at which it is achieved. Still, we see that lowering the actual angle of rotation $\theta$ makes the QFI to increase, approaching the curve we would obtain with pure dephasing (we recall that sending $\theta$ to zero is a way to {annihilate the effect of} the tilting noise). A remarkable phenomenon also appears for $\theta$ below a certain threshold:  there is a crossing between the tails of the pure dephasing noise and the general noise curves. Despite the highest value is always reached by the pure dephasing curve, having a little tilting noise makes the QFI to decrease slower for an increasing number of applications of the gate (see Fig. \ref{fig:decouple}). 
%

{As a second example,  we consider a polar initial state, a qubit with $\alpha \simeq 0$.} In this case, we first notice that the QFI is much smaller than the one obtained for $\alpha=\frac{\pi}{2}$, but this regime is still interesting for the appearance of a  { peculiar metrological behavior}: {if we split $\vec b_0$ in the two components $\vec b_0^{\perp}$ and $\vec b_0^{\shortparallel}$  along the central axis and perpendicular to it, the total QFI is well approximated by the sum of the two QFIs obtained starting with the initial states $\vec b_0^{\perp}$ and $\vec b_0^{\shortparallel}$, resepctively. The first contribution follows the same behavior of the $\alpha=\frac{\pi}{2}$ case. In particular, it is nearly equal to the pure dephasing curve when the tilting noise is very low. The other contribution, instead, can stay much above the pure dephasing curve. For this reason we can find configurations (typically for $\vec b_0^{\shortparallel}>\vec b_0^{\perp}$) where the QFI for general noise is way higher than the pure dephasing one.} This means that in this situations is better to have also a small tilting noise than just the dephasing (see Fig. \ref{fig:decouple}).

\section{Summary and conclusions}
\label{sec:Summary}
{In this paper, we have analyzed the time evolution of a qubit undergoing different noisy quantum  gates with the aim of maximizing the Quantum Fisher Information (QFI) associated to the angle of rotation, i.e. to optimize the characterization of the gates themselves. We have employed  Bloch notation for the state at step $t$, and described the gate as a non unitary quantum operation depending  only on the gate parameter $\theta$. Since every qubit gate can be seen as a rotation, the noise can be seen  as composed of two different sources of fluctuations: deviations from the desired rotation axis (tilting), and fluctuations around the desired 
value of the rotation angle $\theta$ (dephasing). 

Both type of noise have been described as classical fluctuations governed by a Von Mises-Fisher distribution that quantifies the deviation(s) from the desired rotation.  We have initially studied the effect of the two noise contribution separately and found that pure dephasing does not introduce dramatic differences compared to the ideal case. In particular, the QFI of $\theta$ does not depend on $\theta$ itself, and is still maximized selecting a pure initial state that is orthogonal to the rotation axis. For pure tilting, instead, the QFI depends on the parameter, favouring small rotation angles. We have then considered the occurrence of both noises at the same time to describe the most general scenario. In this case, we have found situations in which the presence of tilting is beneficial, as it makes the dephasing noise less detrimental. 

Overall, compared to the noiseless case, in which the QFI grows quadratically with the number of steps, we have found that the presence of noise leads to a non monotonic behavior, and to the appearance of a maximum in the QFI. In other words, there exists an optimal number of steps (i.e., number of gate applications) to be used  in order to precisely characterize the action of the gate.}

\begin{acknowledgments}
This work has been done under the auspices of GNFM-INdAM and has been partially supported by MUR and EU through the projects PRIN22-2022T25TR3-RISQUE and PRIN22-PNRR-P202222WBL-QWEST. The authors thank Andrea Secchi for fruitful discussions about the impact of noise on the QFI.
\end{acknowledgments}
\onecolumngrid
\appendix
\section{Proof that one can consider $\hat z$ rotations only without loss of generality}
\label{app:NoLoss}
Throughout the paper, we often choose to work with $\hat z$ rotations. We show here that the case concerning a general axis can be then obtained in a straightforward way. For standard rotations, a simple change of basis gives this relation
\begin{equation}
\label{ChangeOfBasis}
    R_{\hat n}(\theta)=R^{-1}_{nz} R_{\hat z}(\theta)R_{nz}
\end{equation}
where $R_{nz}$ is the rotation bringing the axis $\hat n$ into the axis $\hat z$: \begin{equation}
\label{NZTransformation}
    \hat z= R_{nz} \hat n.
\end{equation}
Of course it acts in the other way around on rotations, transforming a $\hat z$ rotation in a $\hat n$ rotation.\\
It is trivial to show that this is true also when considering a noisy rotation affected by dephasing: 
\begin{align}
    \Tilde{R}_{\hat n, k_D}(\theta)&=\int_{-\pi}^{\pi} p_{k_T}(\epsilon) R_{\hat n}(\theta+\epsilon) d\epsilon = \int_{-\pi}^{\pi} p_{k_T}(\epsilon) R^{-1}_{nz} R_{\hat z}(\theta+\epsilon)R_{nz} d\epsilon \nonumber \\
    &= R^{-1}_{nz}\left[\int_{-\pi}^{\pi} p_{k_T}(\epsilon) R_{\hat z}(\theta+\epsilon) d\epsilon \right] R_{nz} =R^{-1}_{nz} \Tilde{R}_{\hat z, k_D}(\theta) R_{nz}.
\end{align}
For the tilting case we have that
\begin{equation}
    \Tilde{R}_{\hat z, k_T}(\theta)= \int_{Sphere} p_{k_T}(\hat v \cdot \hat z) R_{\hat v}(\theta) d \hat v
\end{equation}
then 
\begin{equation}
   R^{-1}_{nz} \Tilde{R}_{\hat z, k_T}(\theta)R_{nz}= \int_{Sphere} p_{k_T}(\hat v \cdot \hat z) R^{-1}_{nz}R_{\hat v}(\theta)R_{nz} d \hat v.
\end{equation}
We can choose to integrate in $\hat w$ ($d \hat w=d \hat v$) such that $R^{-1}_{nz}R_{\hat v}(\theta)R_{nz}=R_{\hat w}(\theta)$ and so $\hat w=R^{-1}_{nz} \hat v$. We have then 
\begin{equation}
    \hat v\cdot \hat z=\hat v \cdot R_{nz}R^{-1}_{nz}\hat z= R^{-1}_{nz} \hat v \cdot R_{nz}^{-1} \hat z= \hat w \cdot \hat n,
\end{equation}
and so
\begin{align}
   R^{-1}_{nz} \Tilde{R}_{\hat z, k_T}(\theta)R_{nz}=& \int_{Sphere} p_{k_T}(\hat w \cdot \hat n) R_{\hat w}(\theta) d \hat w = \Tilde{R}_{\hat n, k_T}(\theta).
\end{align}
Given this, it is trivial to show that it is still valid when both tilting and dephasing are considered.
This tells us that the noisy rotations related to a general axis are obtained by simply rotating the ones obtained for the $\hat z$ axis. So, the initial vector will evolve as
\begin{equation}
    \vec b _t= \left[R^{-1}_{nz} \Tilde{R}_{\hat z}(\theta)R_{nz}\right]^t \vec b_0=R^{-1}_{nz}\left[\Tilde{R}_{\hat z}(\theta)\right]^t R_{nz} \vec b_0.
\end{equation}
We can also show that, thanks to rotational symmetry, it suffices to compute the QFI for the $\hat z$ case. Indeed, the QFI after $t$ applications of the noisy gate $\Tilde
R_{\hat n}(\theta)$ on the initial vector $\vec b_0$ is
\begin{align}
    H_{\theta}(t) & = \vert \partial_{\theta} \vec b_t \vert ^2+ \frac{\left(\vec b_t \cdot \partial_{\theta}\vec b_t\right)^2}{1-\vert \vec b_t \vert ^2}  \nonumber \\
&= \vert \partial_{\theta} \Tilde
R_{\hat n}^t(\theta)\vec b_0 \vert ^2+ \frac{\left(\Tilde
R_{\hat n}^t(\theta)\vec b_0 \cdot \partial_{\theta}\Tilde
R_{\hat n}^t(\theta)\vec b_0\right)^2}{1-\vert \Tilde
R_{\hat n}^t(\theta)\vec b_0 \vert ^2} \nonumber \\
& = \vert R^{-1}_{nz} \partial_{\theta}\Tilde{R}_{\hat z}^t(\theta) R_{nz}\vec b_0 \vert ^2+ \frac{\left(R^{-1}_{nz}\Tilde{R}_{\hat z}^t(\theta) R_{nz}\vec b_0 \cdot R^{-1}_{nz}\partial_{\theta}\Tilde{R}_{\hat z}^t(\theta) R_{nz}\vec b_0\right)^2}{1-\vert R^{-1}_{nz}\Tilde{R}_{\hat z}^t(\theta) R_{nz}\vec b_0 \vert ^2} \nonumber \\
& =  \vert \partial_{\theta}\Tilde{R}_{\hat z}^t(\theta) R_{nz}\vec b_0 \vert ^2+ \frac{\left(\Tilde{R}_{\hat z}^t(\theta) R_{nz}\vec b_0 \cdot \partial_{\theta}\Tilde{R}_{\hat z}^t(\theta) R_{nz}\vec b_0\right)^2}{1-\vert \Tilde{R}_{\hat z}^t(\theta) R_{nz}\vec b_0 \vert ^2}
\end{align}
where we used the fact that a rotation such $R^{-1}_{nz}$ does nothing when acting inside a modulus square or on both factors of a scalar product. We can clearly see that it's the same QFI we would have obtained for the $\hat z$ case, but acting on a different initial vector $R_{nz}\vec b_0$. 
\section{Noisy rotation matrices}
\label{app:NoisyMatrices}
Here we show the computation required to obtain the noisy rotation matrices and, just for the dephasing noise, the expression of the QFI.
\subsection{Dephasing noise}
We want to compute the "noisy rotation" in the case in which the noise affects just the angle of rotation, preserving the axis. Following the prescription of eq.(\ref{NoisyRotationGeneral}) we write
\begin{align}
    \Tilde{R}_{k_D}(\theta)=\int_{-\pi}^{\pi} \frac{e^{k_D\cos\epsilon}}{2\pi I_0(k_D)} R_{\hat z}(\theta+\epsilon) \,d\epsilon,
\end{align}
recalling that with $I_j(x)$ we denote the modified Bessel function of order $j$.
Expanding the Von Mises distribution as in eq.(\ref{VMExpansion}) we get
\begin{equation}
    \Tilde{R}_{k_D}(\theta)=\int_{-\pi}^{\pi}\frac1{2\pi} \left[1+\frac2{I_0(k_D)} \sum_{j=1}^{\infty} I_j(k_D) \cos(j\epsilon)\right]
    \left( \begin{array}{ccc}
         \cos(\theta+\epsilon)  & -\sin(\theta+\epsilon) & 0 \\
        \sin (\theta+\epsilon) & \cos(\theta+\epsilon) & 0\\
        0&0&1
    \end{array}\right)d\epsilon
\end{equation}
By expanding also the sines and cosines we will have only integrals of these types
\begin{align*}
    S=&\int_{-\pi}^{\pi}\frac1{2\pi} \biggl[\sin\epsilon+\frac2{I_0(k_D)} \sum_{j=1}^{\infty} I_j(k_D) \cos(j\epsilon)\sin(\epsilon)\biggr] d\epsilon \\
    =&\int_{-\pi}^{\pi}\frac1{2\pi} \biggl\{\sin\epsilon+\frac2{I_0(k_D)} \sum_{j=1}^{\infty} I_j(k_D) \frac12\left[\sin((j+1)\epsilon)-\sin((j-1)\epsilon) \right]\biggr\} d\epsilon 
    =0\\
    C=&\int_{-\pi}^{\pi}\frac1{2\pi} \biggl[\cos\epsilon+\frac2{I_0(k_D)} \sum_{j=1}^{\infty} I_j(k_D) \cos(j\epsilon)\cos(\epsilon)\biggr] d\epsilon \\
    =&\int_{-\pi}^{\pi}\frac1{2\pi} \biggl\{\cos\epsilon+\frac2{I_0(k_D)} \sum_{j=1}^{\infty} I_j(k_D) \frac12  \left[\cos((j+1)\epsilon)+\cos((j-1)\epsilon) \right]\biggr\}d\epsilon
    = \frac{I_1(k_D)}{I_0(k_D)}
\end{align*}
resulting in the pure dephasing noisy matrix
\begin{equation}
    \Tilde{R}_{k_D}(\theta)=\left(\begin{array}{ccc} 
       \frac{I_1(k_D)}{I_0(k_D)}\cos\theta  & -\frac{I_1(k_D)}{I_0(k_D)}\sin\theta & 0 \\
        \frac{I_1(k_D)}{I_0(k_D)}\sin \theta & \frac{I_1(k_D)}{I_0(k_D)}\cos\theta & 0\\
        0&0&1 
    \end{array}\right).
\end{equation}
So, starting with the initial vector $\vec b_0=\vert \vec b_0\vert\left( \sin{\alpha}\cos{\gamma}, \sin{\alpha}\sin{\gamma}, \cos \alpha \right)^T$, after $t$ applications of the noisy gate we are left with
\begin{equation}
    \vec b_t=\vert \vec b_0\vert \left(
    \begin{array}{c}
         \left[\frac{I_1(k_D)}{I_0(k_D)}\right]^t\sin{\alpha}\cos{(t\theta+\gamma)}\\ \left[\frac{I_1(k_D)}{I_0(k_D)}\right]^t\sin{\alpha}\sin{(t\theta+\gamma)}\\
         \cos \alpha
    \end{array}\right)
    \quad
         \mbox{and}
         \quad
         \partial_{\theta}\vec b_t=\vert \vec b_0\vert \left(
         \begin{array}{c}
         \;-t \left[\frac{I_1(k_D)}{I_0(k_D)}\right]^t\sin{\alpha}\sin{(t\theta+\gamma)}\\ t \left[\frac{I_1(k_D)} {I_0(k_D)}\right]^t\sin{\alpha}\cos{(t\theta+\gamma)}\\
         0
    \end{array}\right).
\end{equation}
Similarly to the noiseless case, $\vec b_t \cdot \partial_{\theta} \vec b_t=0$ , and using eq.(\ref{BlochQFI}) we find
\begin{align}
    H_{\theta}(t)=\vert \partial_{\theta}\vec{b_t}\vert^2=\vert \vec b_0\vert^2\left[\frac{I_1(k_D)} {I_0(k_D)}\right]^{2t} t^2 \sin^2 \alpha\,.
\end{align}
\subsection{Tilting noise}
Here we show how to obtain the "noisy rotation" matrix in the case of pure tilting noise. According to eq.(\ref{NoisyRotationGeneral}), we need to integrate a rotation around a random axis with a Von Mises Fisher probability density centered on  $\hat z$. If we write the rotation axis $\hat n$ in polar coordinates  we have
\begin{equation}
\label{Tiltingdef}
    \Tilde{R}_{\hat z, k_T}(\theta)=\int_0^{2\pi}\!\!\!d\varphi\int_0^{\pi}\!\!\!d\phi\,\sin\phi\, R_{\hat n}(\theta)\, \frac{k_T \,e^{k_T \cos \phi}}{4\pi \sinh k_T}    .
\end{equation}
Here $k_T$ is the concentration parameter related to the Von Mises Fisher probability distribution and the 3D rotation can be written as
\begin{equation}
    \label{Rotation_app}
    R_{\hat n}(\theta)= 
        \left(\begin{array}{ccc} \cos\theta+n_x^2(1-\cos\theta) &  n_xn_y(1-\cos\theta)-n_z \sin\theta & n_xn_z(1-\cos\theta)+n_y\sin\theta \\ n_xn_y(1-\cos\theta)+n_z \sin\theta & \cos\theta+n_y^2(1-\cos\theta) & n_yn_z(1-\cos\theta)-n_x \sin\theta \\ n_xn_z(1-\cos\theta)-n_y \sin\theta & n_yn_z(1-\cos\theta)+n_x \sin\theta & \cos\theta+n_z^2(1-\cos\theta)\end{array} \right).
\end{equation}

{By means of the integrals}
\begin{align}
    &\int_0^{\pi}\frac{k_T \sin^3\phi}{4\pi \sinh(k_T)}e^{k_T \cos \phi} d\phi=\frac{k_T\coth k_T-1}{\pi k_T^2}\\
    &\int_0^{\pi}\frac{k_T \sin\phi\cos\phi}{4\pi \sinh(k_T)}e^{k_T \cos \phi} d\phi=\frac{k_T\coth k_T-1}{2\pi k_T}\\
    &\int_0^{\pi}\frac{k_T \sin\phi\cos^2\phi}{4\pi \sinh(k_T)}e^{k_T \cos \phi} d\phi=\frac{k^2_T-2 k_T\coth k_T+2}{2 \pi k_T^2}.
\end{align}
{we can get the noisy rotation for pure tilting}
\begin{align}
\label{TiltingMain_app}
    \Tilde{R}_{\hat z, k_T}(\theta)= 
         \left( \begin{array}{ccc} \cos\theta + A\,(1-\cos\theta) & -B\,\sin\theta  & 0 \\ B\,\sin\theta  & \cos\theta+A\, (1-\cos\theta) & 0 \\ 0 & 0 & \cos\theta+ C\,(1-\cos\theta)\end{array}\right),
\end{align}
where $B=(k_T \coth k_T -1)/k_T$, $A=B/k_T$, $C=(k_T^2-2 k_T \coth k_T +2)/k_T^2$.

\subsection{General noise}
Here instead we assume that noise affects both the axis and the angle of rotation. Similarly to what we did above, we compute the noisy rotation matrix, this time integrating over all the three degrees of freedom of a 3-dimensional rotation.
\begin{equation}
    \Tilde{R}_{\hat z, k_D, k_T}(\theta)= \int_0^{2\pi}\!\!\!d\epsilon \,\frac{e^{k_D \cos\epsilon}}{2\pi I_0(k_D)} \int_0^{2\pi}\!\!\!d\varphi \, \nonumber  \int_0^{\pi}\!\!\! d\phi\, \sin\phi\,\frac{e^{k_T \cos{\phi}}}{4\pi\sinh k_T} R_{\hat n}(\theta+\epsilon) 
\end{equation}
where $k_D$ and $k_T$ are the concentration parameters of the probability distributions relating the dephasing and tilting noise respectively. Performing the integrals one obtains
\begin{align}
    \label{GeneralMain_app}
    \Tilde{R}_{\hat z, k_D, k_T}(\theta)
    = 
    \left( \begin{array}{ccc} \cos\theta \mathcal{F}_{k_D}+ A\,(1-\cos\theta\mathcal{F}_{k_D}) & -B\,\sin\theta \mathcal{F}_{k_D} & 0 \\ B\,\sin\theta \mathcal{F}_{k_D} & \cos\theta\mathcal{F}_{k_D}+A\, (1-\cos\theta\mathcal{F}_{k_D}) & 0 \\ 0 & 0 & \cos\theta\mathcal{F}_{k_D}+ C\,(1-\cos\theta\mathcal{F}_{k_D})\end{array}\right),
\end{align}
where $\mathcal{F}_{k_D}=I_1(k_D)/I_0(k_D)$.
\twocolumngrid
\bibliographystyle{apsrev4-2}
\bibliography{bibnqtd.bib}
\end{document}